\documentclass[sn-mathphys]{sn-jnl}% Math and Physical Sciences Reference Style
\usepackage{graphicx}%
\usepackage{array,multirow}%
\usepackage{amsmath,amssymb,amsfonts}%
\usepackage{amsthm}%
\usepackage{mathrsfs}%
\usepackage[title]{appendix}%
\usepackage{xcolor}%
\usepackage{textcomp}%
\usepackage{manyfoot}%
\usepackage{booktabs}%
\usepackage{algorithm}%
\usepackage{algorithmicx}%
\usepackage{algpseudocode}%
\usepackage{listings}%
\usepackage{natbib}

\usepackage{graphicx,amsmath,amssymb,float}
\usepackage{latexsym,natbib,url,hyperref}
\usepackage{wrapfig,lipsum,booktabs}
\usepackage{caption,subcaption}
\usepackage{multirow,enumitem}
\usepackage{multicol}
\usepackage{natbib}
\usepackage{tabularx} % Add this in your preamble

\theoremstyle{thmstyleone}%
\theoremstyle{thmstyletwo}%

\theoremstyle{thmstylethree}%
\begin{document}

\title[Home Insurance Risks]{A Deep Learning–Copula Framework for Climate-Related Home Insurance Risk}


\author{\fnm{Asim K.} \sur{Dey}}\email{a.dey@ttu.edu}

\affil{\orgdiv{Department of Mathematics and Statistics}, \orgname{Texas Tech University}, \orgaddress{\city{Lubbock}, \postcode{79409}, \state{Texas}, \country{United States}}}

%%==================================%%
%% abstract %%
%%==================================%%

\abstract{Extreme weather events are becoming more common, with severe storms, floods, and prolonged precipitation affecting communities worldwide. These shifts in climate patterns pose a direct threat to the insurance industry, which faces growing exposure to weather-related damages. As claims linked to extreme weather rise, insurance companies need reliable tools to assess future risks. This is not only essential for setting premiums and maintaining solvency but also for supporting broader disaster preparedness and resilience efforts. In this study, we propose a two-step method to examine the impact of precipitation on home insurance claims. Our approach combines the predictive power of deep neural networks with the flexibility of copula-based multivariate analysis, enabling a more detailed understanding of how precipitation patterns relate to claim dynamics. We demonstrate this methodology through a case study of the Canadian Prairies, using data from 2002 to 2011.
}

%\keywords{Extreme rainfall, Block maxima approach, Return level, Probability of extreme, Bootstrap confidence interval}



\maketitle

\section{Introduction}
\label{s:intro}

This paper introduces a new approach for evaluating home insurance risk driven by precipitation events. By integrating statistical modeling with deep learning techniques, the proposed framework enables a more comprehensive assessment of climate-related risks to residential properties while also accounting for the uncertainty arising from different climate model projections.

Indeed, while political debates on climate change become increasingly heated, the insurance sector is one of the industrial frontiers already facing tangible challenges associated with climate dynamics. For instance, only in 2023, there were 142 natural catastrophes worldwide, which 
accumulated to \$108 billion insured losses globally~\citep{SRI_2024}. In recent years, severe floods have become more frequent and have caused significant economic damage. According to the Federal Emergency Management Agency (FEMA)’s National Flood Insurance Program (NFIP) USA, insured losses to significant flood events were  \$11.0 billion in 2012, and \$11.5 billion in 2017~\citep{SOA_2,Dey2016}. In 2022, in Canada, there were 15 extreme weather events, with property and casualty insurance claims ranging from \$35 million to \$1 billion, totaling a record \$3.4 billion in insured losses---most caused by water damage.  In 2023, Canadian insurers paid out another record-high \$3.1 billion to policyholders~\citep{SC_2024}. The Intergovernmental Panel on Climate Change (IPCC) has projected that the severity and frequency of rainfall (and, consequently, floods) will further increase~\citep {IPCC_2021}. Hence, flood-related losses are also likely to rise in the future, jeopardizing both the stability of the insurance industry and the sustainability of the economy as a whole.

Several statistical methods, e.g., Generalized Linear Model (GLM), Autoregressive Integrated Moving Average (ARIMA) model, and Bayesian hierarchical model, have been employed for modeling the number of insurance claims~\citep{Haug:etal:2011,Lyubchich:Gel:2017:insurance,Scheel:etal:2013}. Different machine learning and artificial intelligence (AI) techniques, such as Neural Network (NN), Support Vector Machine (SVM), and Random Forest, have also been successfully utilized for modeling climate change and its impact on insurance risk~ \cite{Wu2011SupportVR,CALDEIRA201562,Kelley2018, dey2021c,Rasp2018, Shortridge2019}.

In this paper, we model and predict precipitation damage-related insurance claims for 2021--2030 using deep neural networks. Our predictions are based on the latest climate projections, and the estimated changes in claim dynamics are attributed to the anticipated climate change. However, clearly, all climate projections are associated with uncertainties. Sources of uncertainties include stochasticity of climate responses, uncertainty about the form and parameters of climate and statistical models, and uncertainty about the future long-term climate scenarios, which tends to increase for higher prediction horizons~\citep{Brown:etal:2013,Forest:etal:2002,Lyubchich:etal:2019:variance,Dey2025_PCP}. Note that addressing some uncertainties is easier than others, and analysis of some uncertainty sources due to, e.g., bias in climate models, already constitutes one of the key research areas in climate science. Hence, forecasts of future claim dynamics may differ according to different climate model projections. In turn, in our analysis of climate-induced risks, the projected precipitations are obtained from six different climate model scenarios. Hence, in any given time period in the future, we have six realizations of insurance risks, each associated with one climate model scenario.

One of the widely used methods to combine multiple model outputs in the ensemble of forecasts is Bayesian Model Averaging (BMA)~\citep{Raftery2005,Sloughter2010, Sansom2017,Barnes2019,Dey_et_al_2024}. 
However, one major challenge in using BMA is determining the weights, and there exists no uniformly accepted BMA weights for climate models. To address this restriction, we propose a new probabilistic ensemble technique, a \textit{parametric copula-based ensemble}, for deriving future insurance claim risks based on a multivariate distribution of multiple climate model projections, where the multivariate distribution is determined by a \textit{copula}~\citep{Skla59}.

The key contributions of our paper can be summarized as follows:
\begin{itemize}

    \item We design a deep neural network architecture to capture and model home insurance risks driven by climate factors.
    
    \item We address the future dynamics of the home insurance claims under multiple climate model projections.      
   
   \item We present a new copula-based method that enables the integration of multiple climate model projections into a unified assessment of climate-induced home insurance risk. Although copulas have been applied in ensemble numerical weather prediction~\citep[see][]{Schefzik2013}, their use in the context of climate model ensembles and broader climate data applications remains unexplored.

\end{itemize}

The remainder of this paper is organized as follows. In  Section~\ref{sec:Data} we describe insurance and climate data employed in our analysis. 
In Section~\ref{sec:Modeling}, we introduce the proposed two-step procedure to model precipitation-induced home insurance claims. 
The results of modeling and forecasting are reported in Section~\ref{sec:results}.  Finally, Section~\ref{sec:Conclusion} concludes the paper.

\section{Data}\label{sec:Data}
We obtain daily precipitation-related home insurance claims data for two middle-sized Canadian cities (City~A and City~B) during the 10-year period of 2002--2011 (control period). The city names are suppressed due to data confidentiality. The number of claims is normalized by the number of homes insured in the cities on each day from 2002 to 2011. The daily precipitation data (mm) for the period of 2002--2011 are obtained from the ERA-Interim reanalysis product~\citep{ERA_Interim} with $0.1^{0}$ spatial grid resolution.

The projected precipitation data for 2021--2030 (scenario period) are obtained from six combinations of downscaled global climate models evaluated under Representative Concentration Pathways RCP~4.5 and RCP~8.5: CanESM2 CanRCM4 4.5,  CanESM2 CanRCM4 8.5, GFDL-ESM2M RegCM4 8.5,   GFDL-ESM2M WRF 8.5, MPI-ESM-LR RegCM4 8.5, HadGEM2-ES RegCM4 8.5~\citep{IPCC_RCP,Scinocca2016, Giorgi2015,Lyubchich:etal:2019:variance}.

\begin{figure*}[!ht]
\subcaptionbox{City A}{\includegraphics[width=0.51\textwidth]{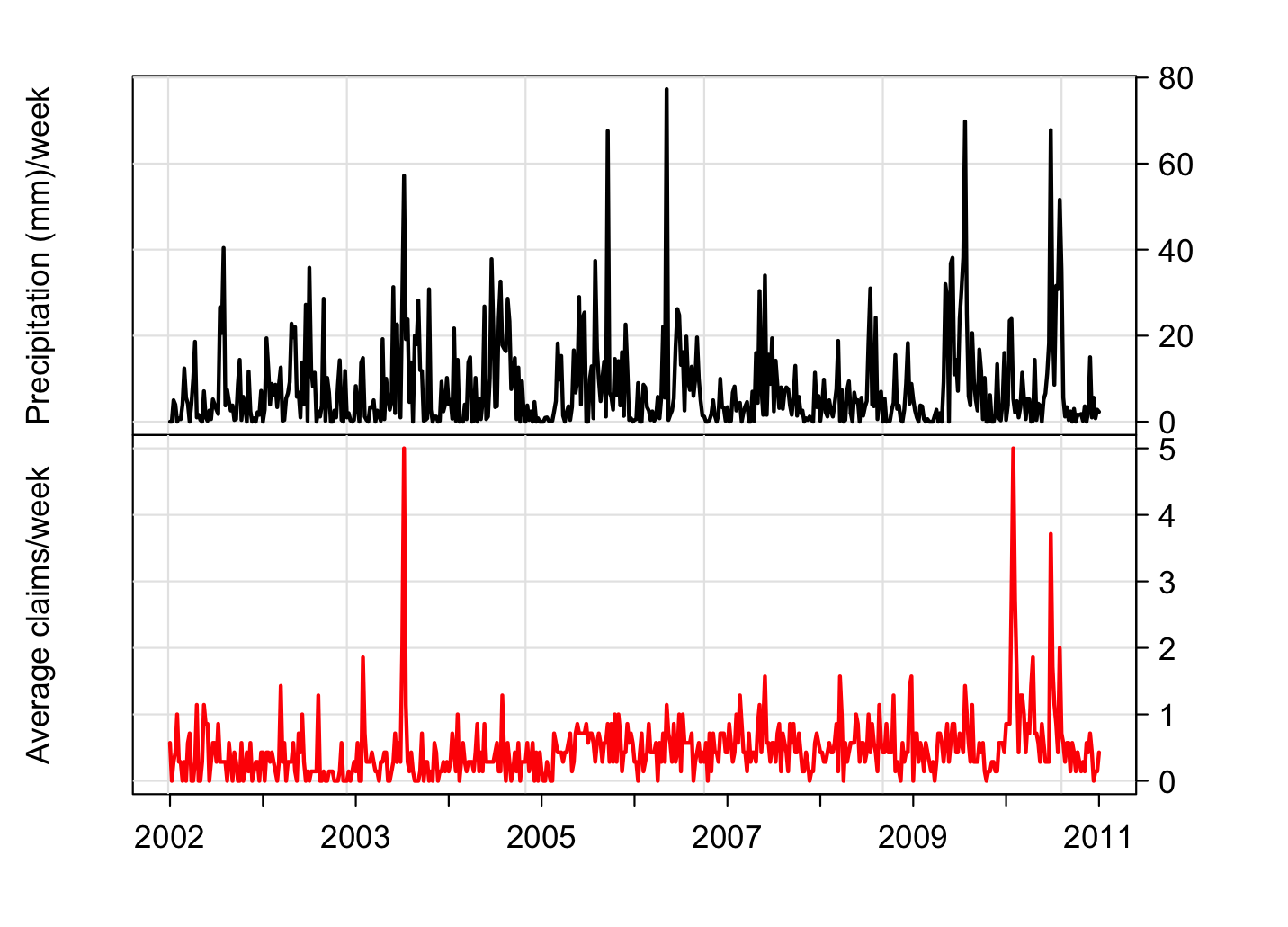}}
\subcaptionbox{City B}{\includegraphics[width=0.51\textwidth]{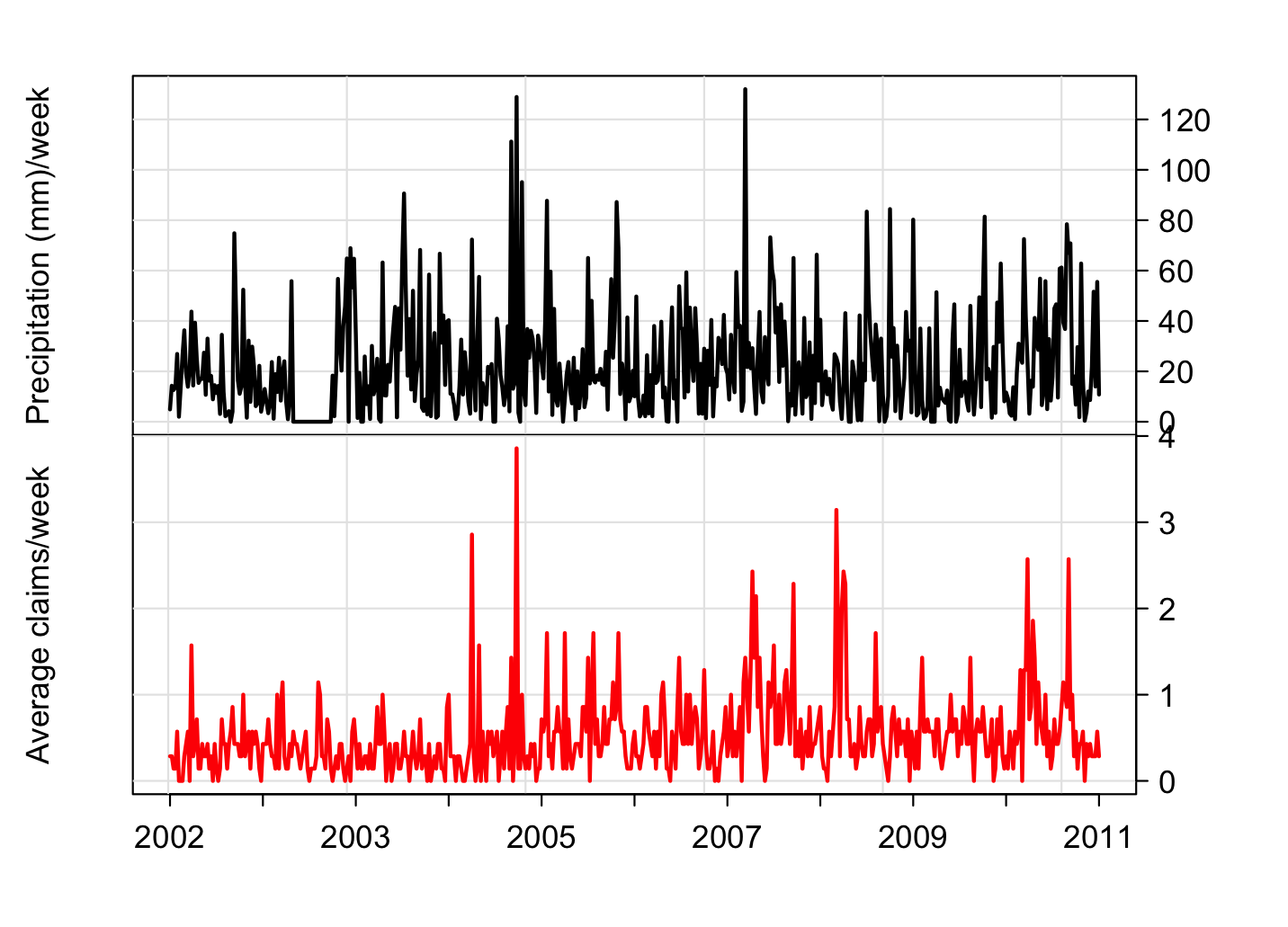}}
\caption{Observed weekly precipitation and number of claims (2002-2011) in City A and City B.}
	\label{figure:timeplot}
\end{figure*}

All daily data are aggregated to the weekly level. Let $N_t$ be the weekly average number of water damage-related insurance claims, and $X_{t}$ be the total precipitation in week $t$. Sometimes the recorded water-related damage in one week is the result of high precipitation in the previous weeks. Hence, we consider an additional explanatory variable $X_{t-k}$ that represents the total precipitation in week $t-k$. We experimentally choose a maximum of $k$ in Section~\ref{sec:results}. We also consider the maximum daily total precipitation in week $t$, $D_t$, and maximum daily total precipitation in the week $t-k$, $D_{t-k}$, as potential covariates. Figure~\ref{figure:timeplot} represents the time plots of the precipitation and number of claims during 2002-2011 in City A and City B. The time series plots do not display clear long-term trends; however, they reveal several sharp, simultaneous spikes across variables, indicating potential interdependence.

%======================

%===========================================

%\newpage
\section{A two-step procedure for quantifying  insurance risk}
\label{sec:Modeling}
We evaluate precipitation-induced home insurance risk using the following two main steps:

\begin{itemize}
    \item \textbf{Modeling} weekly average number of insurance claims using deep neural network based on data in the control period of 2002–2011. \textbf{Prediction} of insurance claims in 2021–2030 based on the trained deep neural network and using projected precipitation from 6 climate models.
    
    \item \textbf{Ensemble} 6 sets of predicted insurance claims based on the multivariate method of copula. The risks of high insurance claim incidents are evaluated by computing the tail probabilities of the derived multivariate distribution.
\end{itemize}

\subsection{Deep neural network models for weekly insurance claims}
\label{sec:models}

We model the weekly average numbers of claims as follows: 
\begin{equation}\label{eq:ncl}
N_t = \varphi(X_t,\cdots,X_{t-k}, D_t,\cdots,D_{t-k}) + \epsilon_{N_t},  
\end{equation}

where $\varphi$ is a function describing the relationships between climate variables and claim dynamics.  We apply a deep neural network to~\eqref{eq:ncl}. The error terms $\epsilon_{N_t}$ are assumed to have zero means (i.e., $E(\epsilon_{Nt})=0$), constant variances, and to be uncorrelated, as in linear regression models. The first term of Eq.~\ref{eq:ncl}, $\varphi(X_t, X_{t-1},\cdots,X_{t-k}, D_t, D_{t-1},\cdots,D_{t-k})$, represents the expected value of $N_t$, i.e., $E(N_t)$.

%==================DNN=========================

Deep neural networks (DNNs) have become one of the most powerful tools in machine learning, especially when it comes to modeling complex and non-linear relationships in data. Unlike traditional linear regression models, which assume a direct, straight-line relationship between inputs and outputs, DNNs are designed to discover and model intricate patterns through multiple layers of abstraction. In other words, a deep neural network (DNN) is a multi-stage regression model, which includes some hidden non-linear layers with multiple hidden units each. The central idea is to extract linear combinations of the inputs as derived features, and then model the target as a nonlinear function of these features based on some activation functions~\citep{goodfellow2016deep}. 

In this study, we develop a deep neural network specifically designed to predict home insurance claims based on precipitation data. The model architecture includes three hidden layers, with each layer containing 64 hidden units. This structure strikes a balance between complexity and efficiency: deep enough to capture non-linearities in the data, but not so deep that it becomes difficult to train or prone to overfitting. The choice of 64 units per layer is based on empirical studies and practical experience, which suggest this size is often sufficient to capture complex features without incurring excessive computational cost~\citep{chollet2018deep}.

The activation function used in each hidden layer is the \textit{rectified linear activation unit} (ReLU). ReLU has become a standard choice in deep learning due to its simplicity and effectiveness. Unlike sigmoid or tanh functions, which tend to saturate and slow down learning in deeper networks, ReLU allows for faster convergence during training. It works by outputting the input directly if it is positive, and zero otherwise. This piecewise linear behavior introduces non-linearity into the model while maintaining efficient computation and helping prevent issues like vanishing gradients. ReLU also encourages sparsity in the network, meaning many neurons are inactive for a given input, which can enhance generalization~\citep{Jarrett_2009,Glorot_2011,maasrectifier_2013}.

While DNNs are powerful, they are also prone to overfitting, especially when trained on limited or noisy data. Overfitting occurs when a model learns to perform well on the training data but fails to generalize to new data. To mitigate this risk, we employ two widely-used regularization techniques: $L_2$ weight regularization and dropout. The $L_2$ regularization, also known as ridge penalty, works by adding a penalty term to the loss function that discourages large weight values. This encourages the network to find simpler solutions and reduces the chance of fitting noise in the training data.

In addition to $L_2$ regularization, we apply dropout, a technique that randomly disables a fraction of the output features from a layer during each training iteration. We use a dropout rate of 0.2, meaning 20\% of the nodes are temporarily ``dropped out'' during each training step. This randomness forces the network to learn redundant representations, making it more robust and less sensitive to specific weights or features. Dropout is one of the most effective ways to prevent overfitting in deep networks, especially when working with relatively small datasets~\citep{tensorflow_2016}.

We can summarize our deep neural network model as follows:
\begin{itemize}
    \item We select a DNN that contains three hidden layers with 64 hidden units in each hidden layer.
    \item We choose the ReLU activation function. The nonlinear ReLU activation function allows the nodes to learn more complex structures in the data and improves the performance of neural networks with piecewise linear hidden units.

    \item To prevent overfitting, we implement the following measures:
      \begin{enumerate}[label=(\alph*)]
          \item We apply $L_2$ weight regularization. 
          \item We apply random dropout of some output features of a layer during training. We choose a dropout rate of 0.2.
      \end{enumerate}

\end{itemize}
%=======================================================================================
\subsection{Ensemble predicted insurance risks}

Using the trained deep neural network described in Section~\ref{sec:models}, we predict insurance claims for the period 2021–2030, based on projected precipitation data from the six climate models outlined in Section~\ref{sec:Data}. From these predictions, we obtain 6 sets of expected weekly number of claims ($\hat{N}_t$) for the scenario period of 2021--2030.

Now we evaluate the tail probabilities of the predicted number of claims in the scenario period to quantify the risk of higher claims in the future. Let $Y_j=[y_{ji}]$, $j=1,2, \cdots, J$; $i=1,2, \cdots, n_w$, be the predicted weekly average claims from the $J=6$ climate projections in 2021--2030, where $n_w$ number of weeks in the scenario period 2021--2030. To capture dependencies among multiple claim predictions resulting from six climate models, we assess the probability of high insurance incidents with a multivariate distribution. Multivariate risk through joint tail probabilities has been studied by~\cite{KRUPSKII2019147, Joe2014}. 
In the actuarial context, \cite{Claudia2012} has proposed a mixed copula approach to study the dependency between the number of insurance claims and average loss. In this study, the claim predictions in the scenario period of 2021–2030 from six different climate projections are studied jointly, using a six-variate distribution~\citep{Moller2009,PINSON201212}. Finally, we treat the tail of the multivariate distribution as a measure of insurance risk. In particular, we evaluate insurance risks based on the following steps:

\begin{itemize}
    \item First, we propose a copula-based ensemble approach to address the dependency among the claim predictions based on multiple climate model outputs.
    \item Then, we quantify the risk of high claim incidents due to adverse atmospheric events through assessment of the tail probability of the derived multivariate distribution.
\end{itemize}

Let the marginal distribution functions of predicted claims from six climate projections, $Y_j$, $j=1,2, \ldots, 6$,  be $F_1(y_1), F_2(y_2), \ldots, F_6(y_6)$, respectively. The joint multivariate distribution of the claims can be written as
\begin{equation}\label{eq:Copula}
F(y_1,\cdots,y_6)=C(F(y_1), \cdots, F_6(y_6)),
\end{equation}
where, $C$ is the copula function~\citep{Skla59}.
In insurance risk modeling, one of the most widely used copula families is the Gumbel copula~\citep{PETERS2014258}. %KRAMER2013
Here we employ a Gumbel copula to combine predictions from the six climate models. For six marginals, the Gumbel copula can be defined as
\begin{equation}\label{eq:copula2}
C(u_1,\cdots,u_6)= \exp\left\{ -\left[   \sum_{i=1}^{6}  \left(-\ln u_i\right)^\theta \right]^{1/\theta}  \right\},
\end{equation}
where $u_i$'s are the dependent cumulative distribution functions, $\theta$ is the copula parameter that measures the degree of dependency among $u_i$'s, $\theta \geqslant 1$ \citep{Nelsen:2006}.

The univariate number of claims can be modeled with different distributions, such as Poisson, negative binomial type I, negative binomial type II, zero-inflated Poisson, and Sichel~\citep{Frees2016}. After studying the goodness of fit (AIC~\citep{Akaike1998} and BIC~\citep{schwarz1978}) of different distributions, we select the negative binomial type I distribution to model claim frequency. The probability that a city has reported $y$ claims ($y = 0, 1,2,\ldots$) can be modeled by the negative binomial Type I distribution as
\begin{equation}\nonumber
 \mathrm{P}(Y=y) = {{\Gamma{(y+  \sigma^{-1})}} \over {\Gamma{\left( \sigma^{-1} \right)}} {\Gamma{(y+1)}}} \left({{\sigma \mu} \over {{1+\sigma \mu}}}\right)^y \left({{1} \over {{1+\sigma \mu}}}\right)^{1/\sigma},
\end{equation}
where $\Gamma(\cdot)$ is gamma function, $\mu > 0$ and $\sigma >0$~\citep{ANSCOMBE1950}.

After selecting appropriate marginal distributions of the predicted number of claims from six climate projections, we model their joint multivariate claim dynamics with the Gumbel copula (Eq.~\ref{eq:Copula} and Eq.~\ref{eq:copula2}).
We define the multivariate tail probability from the fitted  six-variate claims distribution as
\begin{equation}\label{Eq:TailProb}
\Phi(z)=\mathrm{P}(Y_1>z, Y_2>z, \cdots,  Y_6>z),
\end{equation}
where $z$ is a large weekly average insurance claim. Generally, $z$ depends on the primary focus of the risk analysis, e.g., whether the insurance company aims to plan how many assessors need to be hired in the near term, how the spectrum of the insurance products may change in the medium term, and whether building codes shall be revised in the long term.

%================================================================================
\section{Results}
\label{sec:results}
%===============================================================================
We conduct extensive experiments to select the best-performing prediction model for water damage-related weekly number of insurance claims and aggregated losses. We assess the forecasting utility of the precipitation $X_t$, its lags $X_{t-k}$, as well as maximum daily precipitation in that week $D_t$ and its lags $D_{t-k}, k=1,\cdots, 5$. 

The deep neural network models are fitted through the R package \textit{Keras}~\citep{keras3} and \textit{TensorFlow}~~\citep{tensorflow_2016}. We select to showcase the best four of these deep neural network models based on the delivered predictive accuracy (see Table~\ref{table:models}) for City~A and City~B. We consider the root mean squared error (RMSE) as the measure of prediction accuracy~\citep{Harrison2022,HUBER20201420,Akbari2018}.

We find that, for City~A, Model 3, containing information from precipitation, precipitation at lag 1, precipitation at lag 2, and weekly maximum precipitation, delivers the most competitive result. For City~B, Model 4, based on precipitation, precipitation at lag 1, precipitation at lag 2, weekly maximum precipitation, and weekly maximum precipitation at lag 1, yields the most accurate performance. Therefore, we selected DNN Model 3 as our final model for City A and DNN Model 4 as our best model for City B.

\begin{table*}[!ht]
\caption{Model description for the number of claims ($N_t$) with varying predictors, precipitation ($X_t$), precipitation at lag 1 ($X_{t-1}$), precipitation at lag 2 ($X_{t-2}$), weekly maximum precipitation ($D_t$), and weekly maximum precipitation at lag 1 ($D_{t-1}$). The last column represents the root mean squared error (RMSE).}
	\label{table:models}	
	%	\small
	\centering
	\begin{tabular}{c|*{6}{c}r}  
		\hline	
	&	Model &   Predictors  &   RMSE  \\
		\hline

	%	\hline
  &	Model 1 & $X_t$, $X_{t-1}$, $X_{t-2}$ &   0.454 \\
       City A    &	Model 2 & $X_t$, $X_{t-1}$, $D_t$ & 0.463\\

	&	Model 3 & $X_t$, $X_{t-1}$, $X_{t-2}$, $D_t$ &  \textbf{0.453} \\	
    &	Model 4 & $X_t$, $X_{t-1}$, $X_{t-2}$, $D_t$, $D_{t-1}$ & 0.456\\	
		\hline

%	&	Model 1 & $X_1$, $X_2$, $N$ & 33.322 \\
 &	Model 1 & $X_t$, $X_{t-1}$, $X_{t-2}$ & 0.470  \\
       City B    &	Model 2 & $X_t$, $X_{t-1}$, $D_t$ & 0.471 \\

	&	Model 3 & $X_t$, $X_{t-1}$, $X_{t-2}$, $D_t$ & 0.467 \\	
    &	Model 4 & $X_t$, $X_{t-1}$, $X_{t-2}$, $D_t$, $D_{t-1}$ & \textbf{0.461} \\	
		\hline	
	\end{tabular}
\end{table*}

\begin{figure*}[!ht]
%\vspace{-0.51cm}
\centering
	\includegraphics[width=0.95\textwidth]{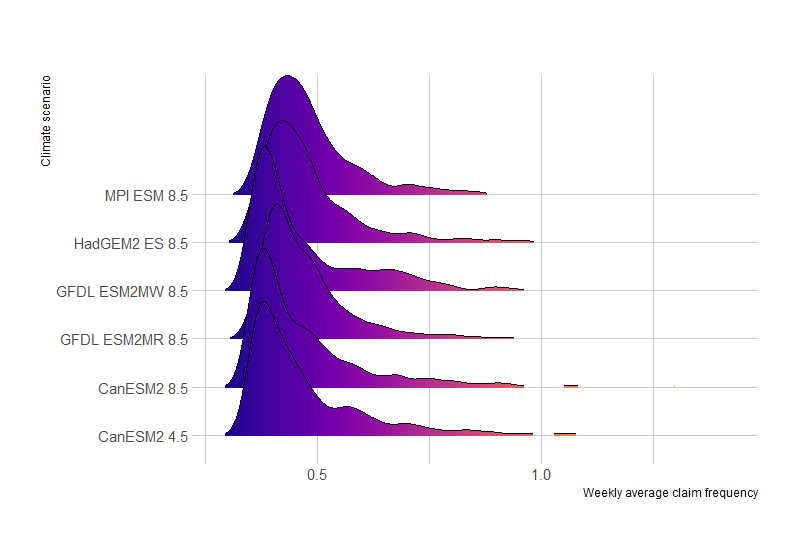}
\caption {City A: Density curves of the predicted weekly average claim frequency in 2021–2030 under different climate models.}
\vspace{-0.2cm}
	\label{fig:Risk_1a}
\end{figure*}

\begin{figure*}[!ht]
%\vspace{-0.51cm}
\centering
    \includegraphics[width=0.95\textwidth]{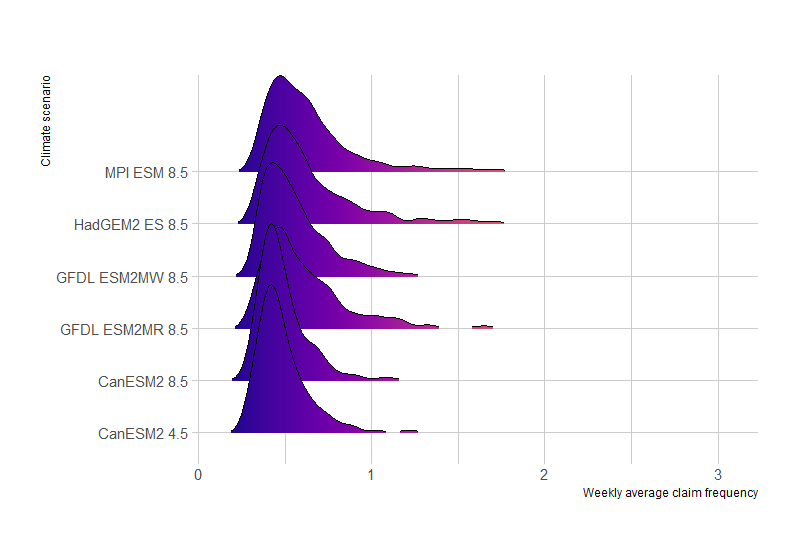}
\caption {City B: Density curves of the predicted weekly average claim frequency in 2021–2030 under different climate models.}
\vspace{-0.2cm}
	\label{fig:Risk_1b}
\end{figure*}

%====================================================================================
Using the selected models, we obtain predictions for the weekly average insurance claims ($\hat{N}_t$) over the scenario period from 2021 to 2030. These predictions are based on future precipitation estimates from six different climate models. Figure~\ref{fig:Risk_1a} and Figure~\ref{fig:Risk_1b} illustrate the resulting probability density functions for the predicted weekly claims in City~A and City~B, respectively. Each distribution reflects the variability introduced by the different climate projections, capturing a range of potential future outcomes. 

A comparison between the two cities reveals a notable difference in the shape of their predicted claim distributions. In particular, City~B exhibits heavier tails, indicating a higher likelihood of extreme values in weekly claims. This suggests that City~B is more susceptible to intense precipitation events that could trigger spikes in claims. In contrast, City~A's distributions appear to have less heavier tails, implying a narrower range of expected outcomes and potentially lower exposure to severe climate-related insurance risks.

We also evaluate 6 sets of predicted claims from the six climate projections in the 10-year scenario period, 2021--2030, with a six-dimensional multivariate distribution. We choose the negative binomial type I distribution to model the marginal distributions of predicted claims during the scenario period, as it provides a better fit compared to alternative distributions. Finally, 
we use the Gumbel copula to combine the six marginals. Table~\ref{tab:Copula_par} 
shows the estimates of the copula parameters with their standard errors for City~A and City~B.

\begin{table*}[!ht]
\caption{Estimates of the Gumbel copula parameter $\theta$ and standard errors.}
\label{tab:Copula_par}
\centering
\begin{tabular}{lcccc} \toprule
&  Estimate ($\hat{\theta}$)   &  Standard error  \\
\hline
City A &  1.327   &  0.024 \\
City B &   1.101    &  0.012  \\
\hline

\end{tabular}
\end{table*}

\begin{figure*}[!ht]
	\centering
    \includegraphics[width=0.49\textwidth]{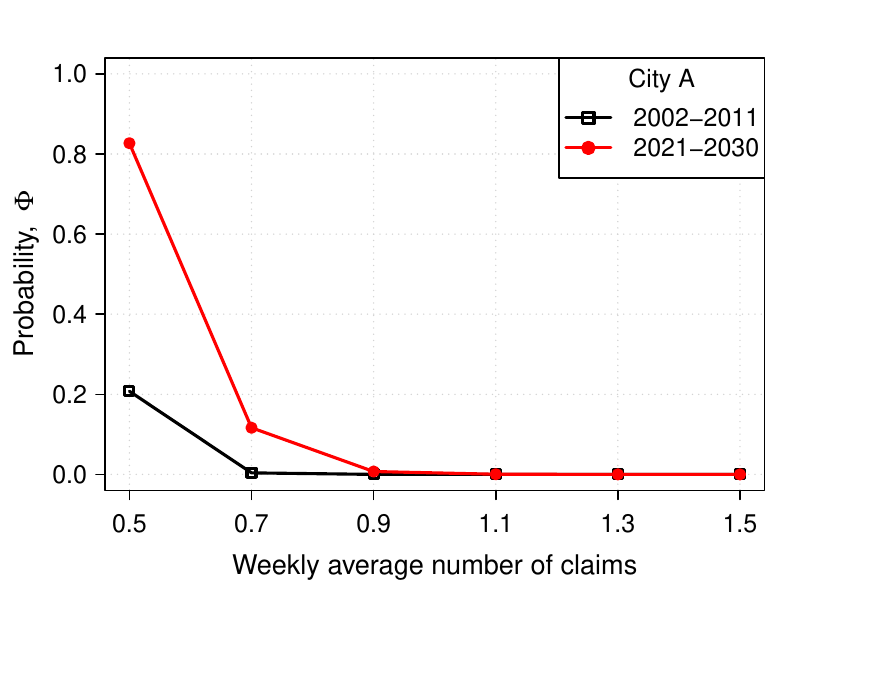}
\includegraphics[width=0.49\textwidth]{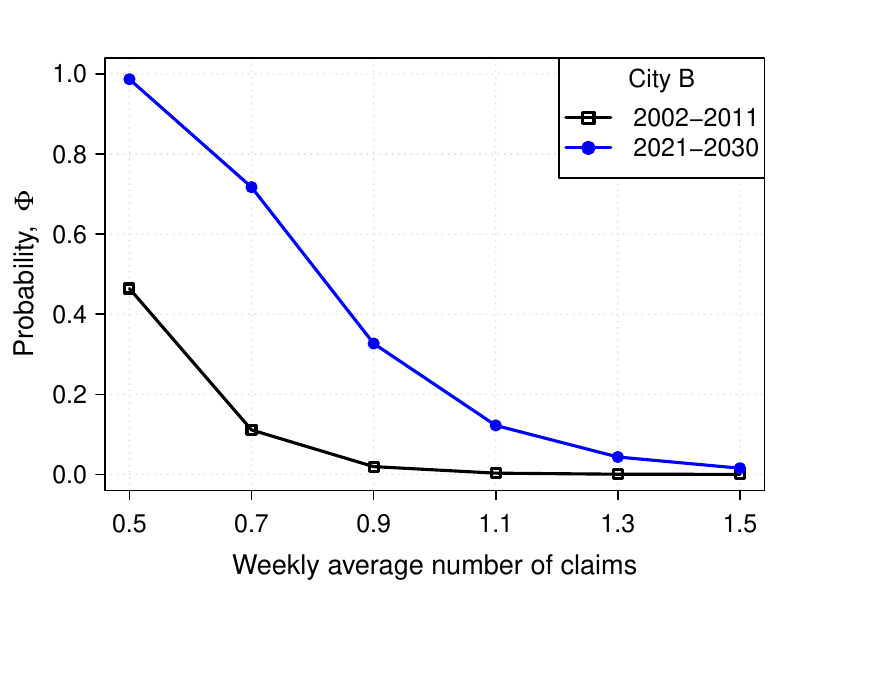}
	\caption {Joint probabilities of high weekly average number of claims in 2021--2030 in City~A and City~B using lognormal marginals and Gumbel copula. }
	\label{fig:RR}
\end{figure*}

We now evaluate the tail of the multivariate distribution using Equation~\ref{Eq:TailProb} as a measure of insurance risk. This tail probability, $\Phi(z)$, captures the likelihood of extreme outcomes, specifically, unusually high average claims, which are critical for understanding worst-case scenarios in insurance planning and pricing. Figure~\ref{fig:RR} presents the computed probabilities of large average claims for both the control period (2002–2011) and the scenario period (2021–2030) in City~A and City~B. The results indicate a noticeable increase in the risk of high average claim occurrences in the future period for both cities. This shift suggests that the potential for extreme insurance losses is projected to grow, likely driven by more frequent or severe precipitation events.

%===========================================================================

\begin{figure*}[!ht]
	\centering
    \includegraphics[width=0.70\textwidth]{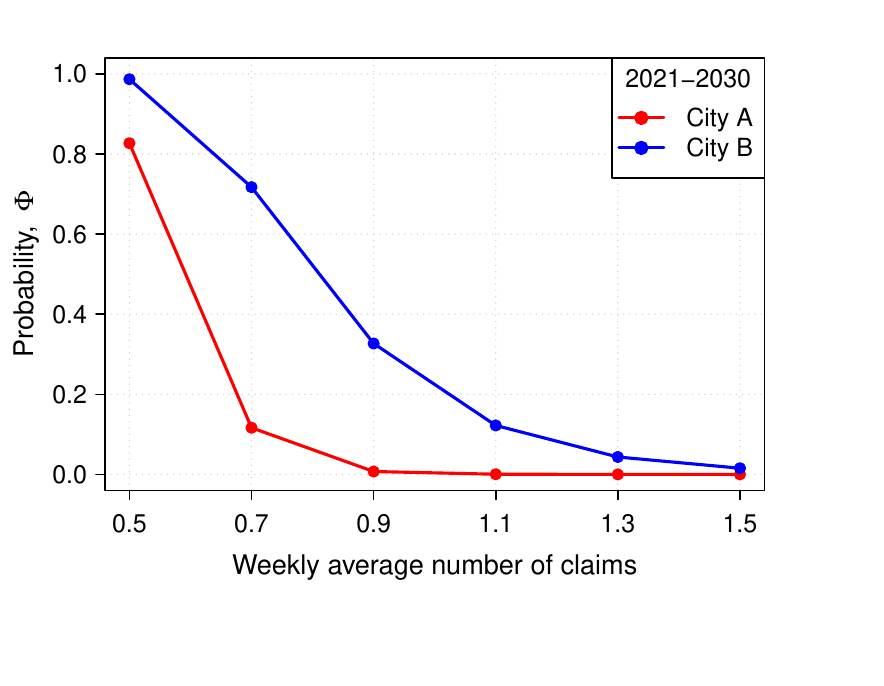}
	\caption {A comparison of the risk of the high weekly average number of claims in City~A and City~B.}
	\label{fig:RR2}
\end{figure*}

We also compare the risk of high insurance claims of City~A to City~B, by evaluating their multivariate tail probabilities $\Phi(z)$ (Figure~\ref{fig:RR2}). We find that City~B has higher risks of a large weekly average number of claims in 2021--2030, compared to City~A. 
Hence, City~B tends to be more vulnerable to climate risk than City~A and is likelier to need the implementation of new city planning policies and risk mitigation strategies. 

%==============================
%\newpage		
\section{Conclusion}
\label{sec:Conclusion}

This study aims to model and forecast the dynamics of climate-induced home insurance claims with respect to changes in precipitation. We employ a deep neural network modeling framework for flexible data-driven characterization of claim dynamics as a function of atmospheric information. We illustrate our approach in application to insurance claims in two middle-sized cities in the Canadian Prairies. Our forecast for the period of 2021-2030 is based on projected precipitation from 6 climate models. We ensemble 6 forecast sets of the average number of insurance claims in 2021-2030 using the multivariate technique of copula. We evaluate the multivariate tails as a measure of insurance risks. Our results indicate that City~B is likely to face substantially higher insurance risks compared to City~A in 2021-2030.

It is important to emphasize that the obtained projections of the future average number of claims account only for changes in future precipitation rather than other meteorological and non-meteorological factors, e.g., seasonal component, socio-economic changes, location and value of assets, that could also influence the dynamics of insurance claims. In the future, we plan to incorporate such information into the assessment of future claim dynamics. We also plan to expand the spatial domain of our analysis by evaluating the dynamics of weather-related home insurance in other cities.

\backmatter

\bmhead{Acknowledgments}
The author thanks Yulia R. Gel of Virginia Tech and Vyacheslav Lyubchich of the University of Maryland Center for Environmental Science for providing data and valuable insights.

%==================
%\bibliographystyle{apalike}
%\bibliographystyle{abbrv}
\bibliography{RefInsurance}

\end{document}